\documentclass[global,twocolumn]{svjour}

\usepackage{epsf}
\usepackage{graphics}

\journalname{Applied Physics B}

\begin{document}

\title{Nonexponential decay of Bose-Einstein condensates: a numerical study
  based on the complex scaling method}

\author{Peter Schlagheck\inst{1} \and Sandro Wimberger\inst{2}}

\institute{Institut f\"ur Theoretische Physik, Universit\"at Regensburg, 93040
  Regensburg, Germany \and Dipartimento di Fisica Enrico Fermi and CNR-INFM,
  Universit\`a degli Studi di Pisa, Largo Pontecorvo 3, 56127 Pisa, Italy}

\maketitle

\begin{abstract}

We study the decay dynamics of an interacting Bose-Einstein condensate in
the presence of a metastable trapping potential from which the condensate can
escape via tunneling through finite barriers.
The time-dependent decay process is reproduced by means of
the instantaneous decay rates of the condensate at a given population of the
quasi-bound state, which are calculated with the method of complex scaling.
Both for the case of a double barrier potential as well as for the case of a
tilted periodic potential, we find pronounced deviations from a
monoexponential decay behaviour which would generally be expected in the 
absence of the atom-atom interaction.

\end{abstract}

\section{Introduction}

With the advent of optical lattices \cite{HecO02JPB,Blo05JPB,MorObe06RMP} and
``atom chips'' \cite{FolO00PRL,FolO02AAMOP}, it became possible to probe
the transport properties of a Bose-Einstein condensate in the mesoscopic
regime.
The unprecedented degree of experimental control in these systems led to
the observation of Bloch oscillations \cite{AndKas98S,MorO01PRL}, the guided
and free propagation of condensates through waveguide structures
\cite{OttO01PRL,HaeO01PRL}, the transport of condensates with ``optical
tweezers'' \cite{GusO02PRL}, as well as the realization of Josephson junctions
\cite{AlbO05PRL} and matter-wave interferometry \cite{SchO05NP}, to mention
just a few examples.
Those experiments typically involve rather small trapping potentials, with
length scales that can be of the order of a few microns.
In such geometries, \emph{decay} mechanisms of the condensate
become a relevant issue.
On the one hand, the condensed state is, at finite atom densities, subject to
depletion, which is caused by the interaction with the thermal cloud and by
three-body collisions.
On the other hand, the condensate can escape from the trapping potential by
tunneling though its barriers, if the chemical potential of the condensed 
atoms exceeds the background potential in the free space outside the trap.
In that case, the self-consistent mean-field state of the condensate is no
longer bound, but rather corresponds to a metastable ``resonance'' state, in a
similar way as, e.g., doubly excited electronic states in the helium atom
\cite{Fri}.

From the theoretical point of view, various methods were used 
\cite{MoiO03JPB,WitMosKor05JPA,CarHolMal05JPB,MoiCed05PRA,SchPau06PRA,WimSchMan06JPB}
to tackle the problem of how to treat ``resonances'', i.e.\ stationary states
that describe the escape of population from an open confinement potential, in
the context of Bose-Einstein condensates.
For linear systems, it is well known that this task is most conveniently
accomplished by applying the method of ``complex scaling'' (or ``complex
rotation'') \cite{BalCom71CMP,Sim73AM,Rei82ARPC,Moi98PR}.
This technique essentially amounts to the complex dilations
$\mathbf{r} \mapsto \mathbf{r} e^{i \theta}$ and $- i \nabla \mapsto - i \nabla e^{-i \theta}$ of the
position and momentum operators in the Hamiltonian that describes the quantum
system under study.
This transformation leads to a nonhermitean Hamiltonian with a complex
eigenvalue spectrum the continuous part of which is rotated to the lower half
of the complex energy plane.
Resonances, i.e.\ decaying states with eigenvalues corresponding to poles of
the resolvent below the real energy axis, are thereby uncovered and can
be calculated using standard diagonalization techniques for complex matrices.
This approach is essentially \emph{exact}, in the sense that no a priori
approximations are introduced in the complex dilation procedure.
Positions and widths of resonances can therefore be calculated with high
precision by means of the complex scaling procedure \cite{Rei82ARPC,Moi98PR}.

The generalization of this approach to Bose-Einstein condensates was recently
accomplished in our previous studies \cite{SchPau06PRA,WimSchMan06JPB}, where
we apply the complex scaling transformation to the nonlinear Gross-Pita\-evskii
equation that describes the mean-field dynamics of the condensate.
In contrast to an alternative approach proposed by Moiseyev and Cederbaum
\cite{MoiCed05PRA}, we explicitly take into account the complex nature of the
wavefunction of the resonance state, which leads to a considerable
complication of the problem due to the resulting nonanalyticity of the
interaction term in the Gross-Pitaevskii equation.
We showed in Ref.~\cite{SchPau06PRA} how this complication can be tackled and
how quasi-bound resonance states of the condensate can be calculated by means
of a real-time propagation approach based on the complex-scaled
Gross-Pitaevskii Hamiltonian.

In this paper, we apply this method in order to calculate specific
time-dependent decay processes of the condensate.
Instead of a direct numerical integration of the time-dependent
Gross-Pitaevskii equation (which is rather consumptive in CPU time for the
evolution time scales under consideration), we compute, with complex scaling,
the decay rates of the quasi-bound state of the condensate at various values
for the effective interaction strength (which would be proportional to the
number of atoms that are populating this quasi-bound state at a given instance 
of time). Then we integrate, on the basis of this information, a
simple rate equation that directly describes the decay of quasi-bound
population.
This approach is comparatively efficient, avoids the introduction of
artificial complex potentials at the grid boundaries in order to absorb the
outgoing population (see, e.g., Ref.~\cite{MoiO03JPB}), and provides physical
insight that could be used to control the decay process in a similar way as
for macroscopic tunneling of condensates in double well potentials
(e.g., \cite{SmeO97PRL,WeiJin05PRA}).

The paper is organized as follows:
In Section \ref{sec:cs}, we establish the general relation between resonance
states of the stationary Gross-Pitaevskii equation and the actual
time-dependent decay process of the condensate.
We furthermore discuss how such resonance states can be calculated by the
method of complex scaling, as was described in more detail in
Ref.~\cite{SchPau06PRA}.
Section \ref{sec:td} contains the numerical results that we obtain for two
paradigmatic examples of metastable confinement configurations:
a harmonic trapping potential with Gaussian envelopes, and a tilted periodic
lattice.
We calculate the time-dependent decay of the condensate, which is, in both
cases, characterized by a pronounced nonexponential nature.

\section{The nonlinear complex scaling approach}

\label{sec:cs}

We consider a Bose-Einstein condensate that is confined within a
cylindrical matter-wave guide with transverse frequency $\omega_\perp$ and evolves in
the presence of a longitudinal potential $V(x)$.
In the ``1D mean-field regime'' \cite{MenStr02PRA} (where the confinement is
strong enough to inhibit transverse excitations within the waveguide, but not
as strong as to enter the Tonks-Girardeau regime \cite{Ols98PRL}), the
dynamics of the condensate is described by the one-dimensional time-dependent
Gross-Pitaevskii equation
\begin{equation}
i \frac{\partial}{\partial t} \psi(x,t) = \left( - \frac{1}{2} \frac{\partial^2}{\partial x^2} + V(x) + g_0
  |\psi(x,t)|^2 \right) \psi(x,t) \label{eq:gp}
\end{equation}
where $x$ denotes the coordinate along the waveguide.
The longitudinal potential $V(x)$ is assumed to provide a local harmonic
confinement with trapping frequency $\omega_\|$, from which the condensate can
escape via tunneling through finite barriers.
Dimensionless variables, defined by setting $\hbar = m = \omega_\| = 1$, are used
throughout this paper.
This means that we express all length scales (including the coordinate $x$) in
units of $a_\| = \sqrt{\hbar / ( m \omega_\|)} $, all energy scales in units of $\hbar \omega_\|$, 
and all time scales in units of $\omega_\|^{-1}$.
The effective one-dimensional interaction strength is, in these units, given
by $g_0 = 2 a_s \omega_\perp$ where $a_s$ denotes the $s$-wave scattering length of the
atoms \cite{Ols98PRL}.

For the description of our theoretical approach, we specifically focus in the
following on the double barrier potential
\begin{equation}
  V(x) = \frac{1}{2} x^2 \exp( - \alpha x^2 ) \label{eq:vdb}
\end{equation}
with $\alpha = 0.1$, which could be experimentally realized, e.g., with red- and
blue-detuned laser beams that are tightly focused onto the waveguide.
Obviously, $V(x)$ does not exhibit any bound state, and the eigenspectrum of
the linear (noninteracting) Hamilton operator is fully continuous.
There exist, however, \emph{quasi-bound} states which are localized in the
well around $x=0$ and which give rise to \emph{resonances} in the energy
spectrum (corresponding to complex poles of the scattering matrix).
In the case of noninteracting atoms ($g_0=0$), such resonance states are
described by time-dependent wavefunctions of the form
$ \psi(x,t) = \psi(x) \exp( - i E t )$, where $\psi(x)$ satisfies the stationary
Schr\"o\-dinger equation for the complex eigenvalue $E = \mu - i \Gamma/2$ and exhibits
outgoing (Siegert) boundary conditions \cite{Sie39PR} 
$\psi(x) \longrightarrow \psi_0 \exp(i k |x|)$, with ${\rm Re}(k) > 0$ for $x \to \pm \infty$.
This latter property expresses the fact that the wavefunction of the decaying
state is characterized by a finite current of atoms that propagate away from
the well.

As a consequence, the atomic population inside the well decays
\emph{exponentially} according to $\propto \exp( - \Gamma t )$ if the system is initially
prepared in the energetically lowest resonance state.
It is quite obvious that this is no longer true in the nonlinear case of
interacting atoms ($g_0 \neq 0$).
There, the tunnel coupling through the barriers explicitly depends, via the
nonlinear term in the Gross-Pitaevskii equation, on the local atomic density,
which in turn induces a temporal variation of the decay rate $\Gamma$.
We therefore naturally obtain, as was also pointed out in
Ref.~\cite{WimO05PRA}, a \emph{nonexponential} decay of the atomic density,
the reproduction of which is the central aim of this paper.

Despite this complication, a description of the decay process of the
interacting condensate in terms of instantaneous quasi-bound states can
nevertheless be justified if the rate $\Gamma$ characterizing the temporal
variation of the density inside the well is rather small compared to the
chemical potential (which should generally be the case if the condensate
escapes via tunneling through finite barriers). 
In such a quasi-stationary situation, we can employ an \emph{adiabatic} ansatz
where the condensate is assumed to remain always in the energetically lowest
(and most stable) resonance state associated with a given instantaneous
density $|\psi(x,t)|^2$.
This resonance state is formally defined, together with its associated complex
eigenvalue $E_g = \mu_g - i \Gamma_g / 2$, by the self-consistent solution of the
nonlinear stationary equation
\begin{equation}
  H(\psi_g) \psi_g(x) = E_g \psi_g(x) \label{eq:gps}\;,
\end{equation}
with
\begin{equation}
  H(\psi) \equiv -\frac{1}{2} \frac{\partial^2}{\partial x^2} + V(x) + g |\psi(x)|^2\;.
  \label{eq:hgp}
\end{equation}
$\psi_g(x)$ is normalized according to the condition
\begin{equation}
  {\mathcal N}[\psi_g] \equiv \int_{-\infty}^\infty |\psi_g(x)|^2 w(x) dx = 1 \, . \label{eq:norm}
\end{equation}
Here $w(x)$ represents a weight function that measures the population inside
the well.
For the double barrier potential (\ref{eq:vdb}), a natural choice for this
weight function would be $w(x) = \theta(a-x) \theta(x+a)$ where $a \equiv 1 / \sqrt{\alpha}$
corresponds to the maximum of the potential and $\theta(x)$ denotes the Heavyside
step function.

In addition, $\psi_g(x)$ should also satisfy outgoing boundary conditions, which
would imply an asymptotic behaviour of the form $\psi_g(x) \propto \exp[ i \int^x k(x')
dx']$ for large $x \to \infty$ (and a similar one for $x \to - \infty$) where the spatial
dependence of the effective wavenumber $k(x)$ accounts for the smooth
variation of the self-consistent potential in Eq.~(\ref{eq:hgp}). 
It was pointed out in Ref.~\cite{SchPau06PRA} that this condition can only be
fulfilled in an approximate way up to a given maximum spatial distance $x_c$
at which the interaction energy $g |\psi_g(x)|^2$ starts to exceed the chemical
potential $\mu$.
For $x > x_c$, the self-consistent quasi-bound state would formally encounter a
singularity, which reflects the fact that explicit time-dependence is expected
beyond that critical distance.

On the basis of these resonance states, we can now formulate the adiabatic
ansatz for $\psi(x,t)$ as
\begin{equation}
  \psi(x,t) = \sqrt{N_0} \psi_{g(t)}(x) \exp\left( - i \int_0^{t} E_{g(t')} dt'
  \right) \, . \label{eq:psi_ad}
\end{equation}
Here, the effective time-dependent interaction strength is given by
$g(t) \equiv g_0 N(t)$ where $N(t)$ denotes the time-dependent population inside
the well, defined by
\begin{equation}
  N(t) \equiv \int_{-\infty}^\infty |\psi(x,t)|^2 w(x) dx \, .
\end{equation}
Using the normalization condition (\ref{eq:norm}) of the resonance state
and taking into account the fact that its eigenvalue $E_g = \mu_g - i \Gamma_g / 2$
is complex, one can straightforwardly derive the implicit expression
\begin{equation}
  N(t) = N_0 \exp\left( - \int_0^t \Gamma_{g(t')} d t' \right) \, . \label{eq:N}
\end{equation}
for the quasi-bound population of the condensate, which can also be formulated
in terms of the ordinary differential equation
\begin{equation}
  \frac{d N}{d t} = - \Gamma_{g(t)} N(t) \label{eq:dN}\;,
\end{equation}
with the initial condition $N(0) = N_0$.
The time-dependent decay process of the condensate can therefore be entirely
reproduced with comparatively little numerical effort if the decay rates
$\Gamma_{g(t)}$ of the instantaneous quasi-bound states $\psi_{g(t)}(x)$ are known.

As was described in detail in Ref.~\cite{SchPau06PRA}, the calculation of the
decay rates can be achieved by the method of \emph{complex scaling}.
This technique essentially amounts to the application of the nonunitary
mapping
\begin{equation}
  \psi(x) \mapsto \psi^{(\theta)}(x) \equiv R_\theta \psi (x) = e^{i \theta / 2 } \psi( x e^{i \theta} ) \label{eq:cs}
\end{equation}
to the wavefunction $\psi(x)$, which corresponds to the complex
dilation $x \mapsto x e^{i \theta}$ of the position operator.
Applying this transformation to the linear stationary Schr\"odinger equation
$H_0 \psi = E \psi$ --- with $H_0$ being defined through Eq.~(\ref{eq:hgp}) via
$H_0 \equiv H(\psi=0)$ --- yields the complex stationary equation
\begin{equation}
  H_0^{(\theta)} \psi^{(\theta)}(x) = E \psi^{(\theta)}(x) \label{eq:sgrot}
\end{equation}
with the complex scaled Hamiltonian
\begin{equation}
  H_0^{(\theta)} \equiv R_\theta H_0 R_\theta^{-1} = - \frac{1}{2} e^{-2 i \theta} \frac{\partial^2}{\partial x^2} +
  V(x e^{i \theta}) \label{eq:h0rot} \, .
\end{equation}

The spectral properties of this nonhermitean Hamiltonian are widely discussed
in the literature on complex scaling
\cite{BalCom71CMP,Sim73AM,Rei82ARPC,Moi98PR}:
While bound states of the original Hamiltonian $H_0$ (which are absent in our
particular case) remain bound after the complex dilation 
(as long as $|\theta| < \pi/4$), the continuum states are ``rotated'' in the complex
energy plane, in such a way that their eigenvalues are located along the axis
$E = \epsilon e^{-2 i \theta}$ with real positive $\epsilon$.
This rotation uncovers the \emph{spectral resonances} of the system, which
correspond to the poles of the analytical continuation of the Green function
$G = (E - H_0 + i \delta)^{-1}$ to the lower half part of the complex energy plane.
Those resonances turn into discrete complex eigenvalues $E_n = \mu_n - i \Gamma_n / 2$ 
under complex dilation, and are represented by normalizable eigenfunctions
that can be straightforwardly calculated by diagonalizing $H_0^{(\theta)}$ in any
numerical basis.

The generalization of this approach to the nonlinear case would be
comparatively straightforward if the replacement $|\psi(x)|^2 \to [\psi(x)]^2$ in the
nonlinear Gross-Pitaevskii Hamiltonian (\ref{eq:hgp}) could be justified.
For this particular case, the implementation of the complex scaling technique
was explained in detail in Ref.~\cite{MoiCed05PRA}.
In reality, however, the wavefunction of the resonance state is
\emph{intrinsically complex} due to the outgoing boundary conditions
(i.e, $\psi(x) \propto \exp(i k |x|)$ for $|x| \to \infty$), and the resulting nonanalyticity
in the Hamiltonian (\ref{eq:hgp}) introduces a major complication of the
problem.
Formally, a second analytic wavefunction $\overline{\psi}$ needs to be
introduced, which coincides with the complex conjugate of $\psi$ on the real axis,
i.e.
\begin{equation}
  \overline{\psi}(x) \equiv \psi^*(x) \qquad \mbox{for real $x$,} \label{eq:psibar}
\end{equation}
and which is independently transformed under the nonunitary dilation operator
$R_\theta$, i.e.
\begin{equation}
  \overline{\psi}(x) \mapsto  \overline{\psi}^{(\theta)}(x) \equiv 
  R_\theta \overline{\psi} (x) = e^{i \theta / 2 } \overline{\psi}( x e^{i \theta} ) \, . \label{eq:csbar}
\end{equation}
The analytic continuation of the stationary Gross-Pitaevskii equation
to the complex domain yields then 
\begin{equation}
  H^{(\theta)}(\psi) \psi^{(\theta)}(x) = E \psi^{(\theta)}(x) \label{eq:gprot}
\end{equation}
where the complex scaled nonlinear Hamiltonian is given by
\begin{equation}
  H^{(\theta)}(\psi) = H_0^{(\theta)} + g e^{- i \theta} \overline{\psi}^{(\theta)}(x) \psi^{(\theta)}(x) \, .
  \label{eq:hrot} 
\end{equation}

The lowest resonance
state of the condensate can be calculated by a real-time propagation approach 
\cite{SchPau06PRA}, i.e.\ by numerically propagating $\psi^{(\theta)}$ 
under the \emph{time-dependent} Gross-Pita\-evskii equation 
\begin{equation}
  i \frac{\partial}{\partial \tau} \psi_\tau^{(\theta)}(x) = H^{(\theta)}(\psi_\tau) \psi_\tau^{(\theta)}(x) \label{eq:gptrot}
\end{equation}
in the complex scaled system, where $\tau$ represents a fictitious numerical
``time'' parameter (which is unrelated to the physical time evolution in the
actual decay process).
In practice, $\psi_\tau^{(\theta)}(x)$ is expanded on a spatial grid, and an implicit
finite-difference scheme is employed to carry out the mapping 
$\psi_\tau^{(\theta)} \mapsto \psi_{\tau+ \delta \tau}^{(\theta)}$ for small time steps $\delta \tau$.
If $\psi_\tau^{(\theta)}$ is renormalized after each propagation step in order to satisfy the
condition (\ref{eq:norm}), the integration of Eq.~(\ref{eq:gptrot})
necessarily converges to the most stable resonance state of the complex
scaled Hamiltonian, which corresponds to the quasi-bound state with the
smallest decay rate \cite{remark}.

The major numerical difficulty in this approach lies in the evaluation of the
nonlinear term in the complex scaled Hamiltonian (\ref{eq:hrot}).
Indeed, $\overline{\psi}^{(\theta)}(x)$ is {\em not} identical to $[{\psi^{(\theta)}}(x)]^*$,
the complex conjugate of $\psi^{(\theta)}(x)$, and needs to be evaluated according to
the relation
\begin{equation}
  \overline{\psi}^{(\theta)}(x) = R_\theta \left( \overline{ R_{-\theta} \psi^{(\theta)} } \right) (x)
  \label{eq:psibarpsi}
\end{equation}
which requires explicit back- and forward-rotations of the complex scaled
condensate wavefunction.
In practice, these rotations (which are also used to evaluate the
normalization condition (\ref{eq:norm})) are numerically performed by
mapping the grid representation of the wavefunction into a nonorthogonal set
of analytic Gaussian orbitals $\phi_\nu(x)$ that are centred around different
positions along the grid, and by using a transformation matrix that contains
the overlap integrals $\int \phi_\nu(xe^{i\theta}) \phi_{\nu'}(x) dx$ as elements (see
Ref.~\cite{SchPau06PRA} for more details).
Such an operation, however, is known to be potentially unstable
\cite{BucGreDel94JPB} and requires great care in the numerical implementation.
It is therefore not obvious to which extent unlimited precision in the decay
rates of the quasi-bound states can be achieved within this nonlinear complex
scaling approach.

\section{Calculation of time-dependent decay processes}

\label{sec:td}

\subsection{Double barrier potential}

\label{sec:db}

Despite this latter complication, we find that the chemical potentials and
decay rates of the self-consistent quasi-bound states of the double barrier
potential (\ref{eq:vdb}) can be calculated in this way with rather good
accuracy, even in case of very strong nonlinearities where the resonance level
lies close the barrier height of $V(x)$.
This was explicitly verified in Ref.~\cite{SchPau06PRA} by comparing the 
resulting values for $\mu_g$ and $\Gamma_g$ with the ones that are obtained from an
alternative approach, which was based on the real-time propagation of the
original (i.e., unscaled) Gross-Pitaevskii equation in the presence of
absorbing boundaries.
Good agreement was generally found between the two approaches 
\cite{SchPau06PRA}.

\begin{figure}[t]
\begin{center}
\epsfxsize\linewidth
\epsfbox{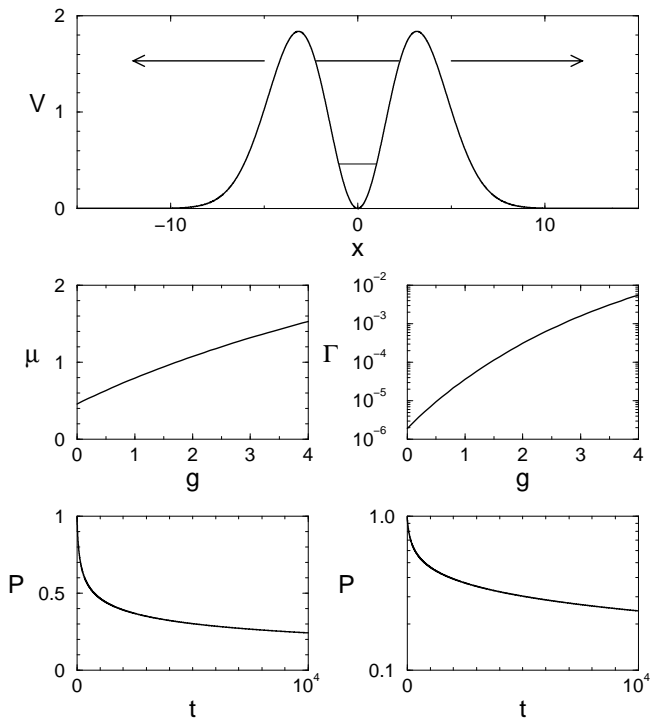}
\caption{
  Decay of a Bose-Einstein condensate in the double barrier potential.
  The upper panel shows the potential (\ref{eq:vdb}) together with the
  chemical potential of the initial quasi-bound state of the condensate 
  (thick horizontal line), which corresponds to the initial value 
  $g(t=0) \equiv g_0 N(t=0) = 4$ of the effective interaction strength.
  During the decay process, the chemical potential decreases with time, due to
  the reduced interaction energy, and approaches the level of the
  noninteracting quasi-bound state (thin horizontal line in the upper panel).
  This decrease of $\mu$ is accompanied by a strong reduction of the decay rate
  $\Gamma$, which can be seen in the two middle panels where $\mu$ and $\Gamma$ are
  plotted as a function of $g$.
  As a consequence, a pronouncedly \emph{nonexponential} decay for the
  quasi-bound population $P(t) \equiv N(t) / N(t=0)$ is obtained from integrating
  the equation $dN/dt = - \Gamma(N) N$, which is displayed in the two lower panels
  (where $P$ is plotted vs.\ $t$ on a linear and logarithmic scale,
  respectively).
  \label{fg:db}
}
\end{center}
\end{figure}

The chemical potentials and decay rates of the lowest resonance state are
plotted in Fig.~\ref{fg:db} as a function of the effective interaction
strength $g$. 
Quite intuitively $\mu_g$ increases with increasing $g$ due to the presence of
the mean-field interaction energy in the nonlinear Gross-Pitaevskii
Hamiltonian.
This increase of the chemical potential results in a dramatic enhancement
of the decay rate $\Gamma_g$, which can be explained by the fact that the effective
imaginary action integral that semiclassically determines the tunneling rate
through the barriers is appreciably reduced with increasing energy.
It was pointed out in Refs.~\cite{MoiO03JPB,CarHolMal05JPB,SchPau06PRA} that
an attractive interaction between the atoms leads to a \emph{stabilization} of
the resonance state, i.e., to a reduction of the chemical potential to values
below $\mu = 0$ where the associated decay rate would vanish.
For the double barrier potential under consideration, this stabilization
process would occur at $g \simeq - 1.1$.

With this information, we can now quantitatively reproduce the time-dependent
decay process of the condensate by means of the integration of the rate
equation (\ref{eq:dN}).
For this purpose, we use the values of the decay rates that are calculated
with the complex scaling method at the equidistant interaction strengths 
$g = 0, 0.1, 0.2 \ldots$, and employ a cubic interpolation to obtain intermediate
values of $\Gamma$.
As initial value of the effective interaction strength, we consider 
$g(t=0) = 4$, where the chemical potential of the quasi-bound state lies
already rather close to the barrier height of the potential.
In the specific case of a condensate of $^{87}$Rb atoms that encounters the
longitudinal and transverse confinement frequencies $\omega_\| = \omega_\perp = 2 \pi \times 10^3$
Hz, this would imply that about $N(0) = 4 / g_0 \simeq 100$ atoms are initially
localized in the single well \cite{note}.

The result of the integration is displayed in the two lower panels of
Fig.~\ref{fg:db}.
We see a clearly nonexponential decay of the bound population,
which reflects the fact that the decay rate decreases with descreasing
interaction strength $g$.
As a consequence, a rather large number of atoms leave the trap during the
first $1000$ units of the evolution time, while the remaining part of the
condensate becomes stabilized and decays, for asymptotically large times, with
the rate $\Gamma_0 \simeq 2 \times 10^{-6}$ of the noninteracting quasi-bound state.
In the example of a $^{87}$Rb condensate in a confinement with trapping
frequency $\omega_\| = 2 \pi \times 10^3$ Hz, the above characteristic time scale of the
nonexponential behaviour would correspond to $t \sim 100$ ms which is of
experimental relevance.

\subsection{Tilted periodic potential}

\label{sec:ws}

Nonexponential features can also be observed in the presence of relatively
\emph{small} interaction strengths, namely if the confinement potential
permits the possibility of \emph{resonant tunneling}.
This is, for instance, the case for the tilted periodic potential
\begin{equation}
  V(x) = \sin^2(x/2) + F x \label{eq:vws}
\end{equation}
that is experimentally realized with optical lattices
\cite{HecO02JPB,Blo05JPB,MorObe06RMP} or, within the atom chip context, by
means of periodic sequences of microfabricated wires \cite{GueO05PRL}.
In this nonlinear Wannier-Stark system, the possibility of resonant tunneling
arises if the local ground state of one of the wells is nearly degenerate with
the first excited state of the adjacent well.
As was pointed out in Ref.~\cite{WimO05PRA}, this near-degeneracy would give
rise to a significant enhancement of the condensate's decay rate.

As in our previous study \cite{WimSchMan06JPB}, we assume that the condensate
is initially confined within one single well of the lattice.
The method of complex scaling can again be used to calculate the chemical
potential and decay rate of the self-consistent quasi-bound state in the
presence of the interaction, even though the tilted potential (\ref{eq:vws})
leads to an asymptotic spatial behaviour of the continuum states that is
substantially different from the previous double-barrier problem.
As was described in detail in Ref.~\cite{WimSchMan06JPB}, additional
complications arise in this potential (such as the existence of many different
self-consistent resonance states with identical decay rates) and technical
modifications need to be implemented in order to achieve good convergence of
the real-time propagation method.
In analogy with the double barrier potential, the weight function that
characterizes the bound population inside the well according to
Eq.~(\ref{eq:norm}) is given by $w(x) = \theta(\pi-x) \theta(x+\pi)$.

\begin{figure}[t]
\begin{center}
\epsfxsize\linewidth
\epsfbox{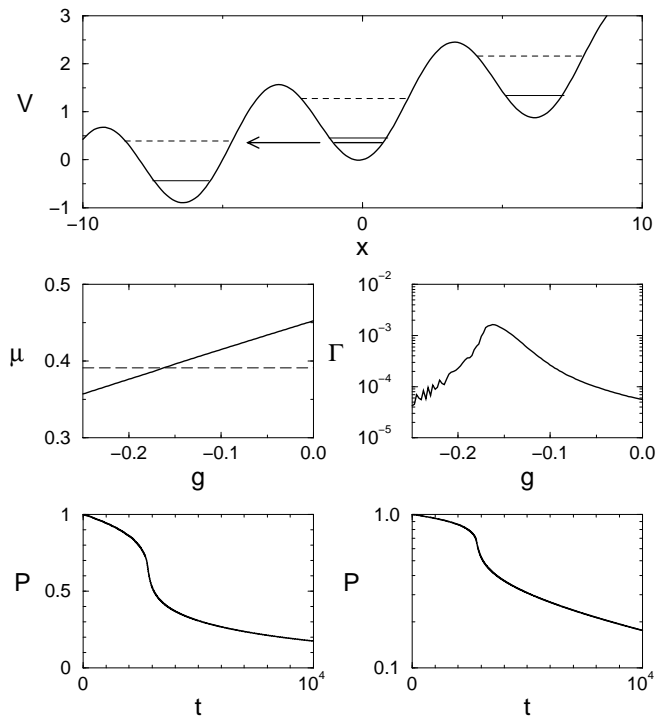}
\caption{
  Decay of a Bose-Einstein condensate in the tilted periodic potential
  (\ref{eq:vws}) with $F = 0.1412$.
  The atomic cloud is assumed to be entirely localized within a single well
  of the lattice, and decays via tunneling through the barrier on the
  left-hand side of the well.
  This decay process can be substantially accelerated in the case of
  \emph{resonant tunneling}, i.e., if the chemical potential of the condensate
  matches the energy of the first excited (noninteracting) state of the
  adjacent well (dashed horizontal lines; the thin solid lines mark the
  noninteracting ground levels of the wells).
  In contrast to the case of the double barrier potential, we consider here a
  relatively weak and attractive initial interaction strength, $g(0) = -0.25$,
  which leads to an initial chemical potential that lies slightly below the
  level of this excited state (thick horizontal line in the upper panel).
  The time evolution of the quasi-bound population $P(t) = N(t) / N(0)$,
  shown in the lower panels, clearly displays the characteristic signature of
  an intermediate resonant tunneling process.
  Note that the wiggles in the decay rates at large $|g|$ (middle right
  panel), which arise from numerical inaccuracies, do not leave significant
  traces in the time evolution of $P(t)$.
  \label{fg:ws}
}
\end{center}
\end{figure}

The resonance-enhanced decay process of the condensate in this tilted lattice
is displayed in Fig.~\ref{fg:ws}, for the tilt strength $F = 0.1412$ at which
the level of the noninteracting local ground state in each well lies slightly
above the level of the first excited state in the adjacent well on the
left-hand side. 
As initial value for the nonlinearity, we consider $g(t=0) = -0.25$, which
would correspond to an \emph{attractive} interaction between the atoms
(which could be realized, e.g., by using condensates with $^7$Li atoms or by
applying Feshbach tuning techniques \cite{PS2002}).
This attractive nonlinearity lowers the chemical potential of the
self-consistent quasi-bound state, in such a way that is becomes shifted below
the level of the first excited (noninteracting) state in the adjacent well.
During the time evolution, the loss of population leads to an increase of
the chemical potential, which, at about $t \simeq 2500$ time units, approaches the
resonance.
At that point, the decay rate becomes drastically increased resulting in a
significant escape of atoms from the well.
As a consequence, the chemical potential of the quasi-bound state quickly
moves out of resonance, and the decay rate becomes again reduced.

The time evolution of the condensate was again calculated by integrating the
rate equation (\ref{eq:dN}), using decay rates that were calculated with the
complex scaling method at the equidistant values $g = 0$, $-0.0025$, $-
0.005$, \ldots of the interaction strength.
The accuracy with which these decay rates could be calculated was not as good
as in the case of the double barrier potential \cite{SchPau06PRA}, which is
clearly reflected by the appearance of wiggles on the left-hand side in the
middle right panel of Fig.~\ref{fg:ws}.
These artificial fluctuations of $\Gamma$, however, leave no significant traces in
the actual time-dependent decay process, as is clearly seen in the lower two
panels of Fig.~\ref{fg:ws}.

This intermediate ``burst'' of atoms should be readily observable within
existing experimental setups based on optical lattices or atom chips.
By imposing a rather weak transverse confinement, the one-dimensional
interaction strength (given by $g_0 = 2 a_s \hbar \omega_\perp$ \cite{Ols98PRL}) can be
appreciable reduced, what should allow one to suppress effects beyond the
mean-field description of the condensate, and additional longitudinal
potentials can, as in the experiment on tunneling well potential
\cite{AnkO05PRL}, be employed to prepare the condensate in one single well
of the lattice.
Obviously, the attractive interaction between the atoms is not a necessary
condition for this nonexponential decay phenomenon:
Indeed, the same effect could be induced with a \emph{repulsively} interacting
species at slightly weaker tilt strengths $F$, where the noninteracting ground
state of the well lies slightly \emph{below} the first excited level of the
adjacent well.
In both cases of attractive and repulsive interaction, the intermediate
enhancement of the decay rate should clearly manifest in the time-of-flight
image of the condensate after the decay process, which would display a
pronounced peak due to the effect of resonant tunneling.

\section{Conclusion}

In summary, we studied the time-dependent decay of Bose-Einstein condensates
in mesoscopic trapping potentials that permit escape by tunneling through
finite barriers.
The decay process of the condensate was reproduced by integrating a simple
rate equation for the quasi-bound population, using instantaneous decay rates
that were computed by means of the method of complex scaling.
This approach is rather efficient as compared to a direct numerical
integration of the time-dependent Gross-Pitaevskii equation, and provides
additional insight into the mechanisms that underly the decay of the
condensate.
Though only applied for one-dimensional configurations, the complex scaling
approach can be straightforwardly generalized to three-dimensional decay
problems, and might furthermore represent a convenient conceptual framework
for treating resonances of the nonlinear Gross-Pitaevskii equation from the
mathematical point of view.

With this approach, we calculated the decay of a Bose-Einstein condensate in a
double barrier potential \cite{MoiO03JPB,CarHolMal05JPB,SchPau06PRA} as well
as in a tilted periodic potential \cite{WimO05PRA,WimSchMan06JPB}.
For this latter case, we find a strong intermediate enhancement of the
tunneling rate, which arises due to a near-degeneracy with a quasi-bound state
in another well of the periodic potential.
This enhancement leads to a pronounced deviation from an exponential behaviour
of the condensate's escape dynamics, which could be controlled by suitable
time-dependent variations of the tilt field, in a similar way as for
macroscopic tunneling in double well potentials \cite{SmeO97PRL,WeiJin05PRA}
and for nonlinear resonant transport through atomic quantum dots
\cite{PauRicSch05PRL}.
Such nonexponential effects should be readily observable in present-day
state-of-the-art experiments on interacting matter waves in mesoscopic
trapping potentials 
\cite{MorObe06RMP,AlbO05PRL,SchO05NP,GueO05PRL,AnkO05PRL,GHF2006}.


\begin{thebibliography}{10}
\bibitem{note}
The number of atoms in the well can be enhanced by choosing a weaker
transverse confinement, e.g.\ $\omega_\perp = 10^2$ Hz for which on would obtain $N(0)
\simeq 1000$ atoms. In that case, however, three-dimensional calculations ought to
be performed in order to make reliable quantitative predictions for the decay
process.
\end{thebibliography}
\end{document}